\documentclass[runningheads]{llncs}

\usepackage{graphicx}
\usepackage{algorithm}
\usepackage{algpseudocode} % ADDED by CCD, for algorithms
\usepackage{mathrsfs}
\usepackage{textcomp}
\usepackage{xcolor}
\usepackage{multirow}
\usepackage{tipa}
\usepackage{forest}
\usepackage{subfig}
\usepackage[normalem]{ulem}

\usepackage{booktabs}

\begin{document}
	
	\title{Bilingual Speech Recognition by Estimating Speaker Geometry from Video Data\thanks{This material is based upon work supported by the National Science Foundation under Grant No.1613637, No.1842220, and No.1949230.}}

	\titlerunning{Bilingual Speech Recognition by Estimating Speaker Geometry}
	\author{Luis Sanchez Tapia\inst{1} \and
		Antonio Gomez\inst{1} \and 
		Mario Esparza\inst{1} \and
		Venkatesh Jatla\inst{1} \and
		Marios Pattichis\inst{1} \and
		Sylvia Celed\'{o}n-Pattichis \inst{2} \and
		Carlos L\'{o}pezLeiva \inst{2}
	}

	\authorrunning{Sanchez, L. et al.}
	
	\institute{Department of Electrical and Computer Engineering \\ 
		The University of New Mexico, Albuquerque, NM, USA. \\
		\email{\{luis2sancheztapia, agsuper, javesparza, venkatesh369, pattichi\}@unm.edu}
		\and
		Department of Language, Literacy, and Sociocultural Studies\\
		The University of New Mexico, Albuquerque, NM, USA. \\
		\email {\{sceledon, callopez\}@unm.edu}
	}

	\maketitle
	
	\begin{abstract}
		Speech recognition is very challenging
		in student learning environments that
		are characterized by significant cross-talk
		and background noise.
		To address this problem,	
		we present a bilingual speech recognition system
		that uses an interactive video analysis system
		to estimate the 3D speaker geometry 
		for realistic audio simulations.
		We demonstrate the use of our system
		in generating a complex audio dataset
		that contains significant cross-talk and
		background noise that approximate
		real-life classroom recordings.
		We then test our proposed system
		with real-life recordings.   
		
		In terms of the distance of the speakers from the 
		microphone, our interactive video analysis system
		obtained a better average error rate of 10.83\% 
		compared to 33.12\% for a baseline approach.
		Our proposed system gave an accuracy
		of 27.92\% that is 1.5\% better than Google
		Speech-to-text on the same dataset.
		% Explain what is sensitivity|specificity?
		In terms of 9 important
		keywords, our approach gave an average sensitivity
		of 38\% compared to 24\% for Google Speech-to-text,
		while both methods maintained high average specificity 
		of 90\% and 92\%.

		On average,
		sensitivity improved from 24\% to 38\% for our proposed approach.
		On the other hand, 
		specificity remained high for both methods (90\% to 92\%).

	\end{abstract}
	
	\section{Introduction}
	% word video should appear in the first line.
	Human activity recognition can strongly benefit from the combined use of audio and video data.
	More recently, audio processing has been used to identify visual events 
	\cite{looking}, \cite{multisensory}.
	For our paper, we want to investigate the use of video data to reconstruct the 
	speaker geometry in 3D and then use this information to develop 
	a speaker recognition system.
	Our approach addresses the strong need to develop 
	a speech recognition system that can
	help transcribe student conversations 
	from video recordings of collaborative learning environments.
	
	\begin{figure}[!t]
		\centering
		\includegraphics[width=\textwidth]{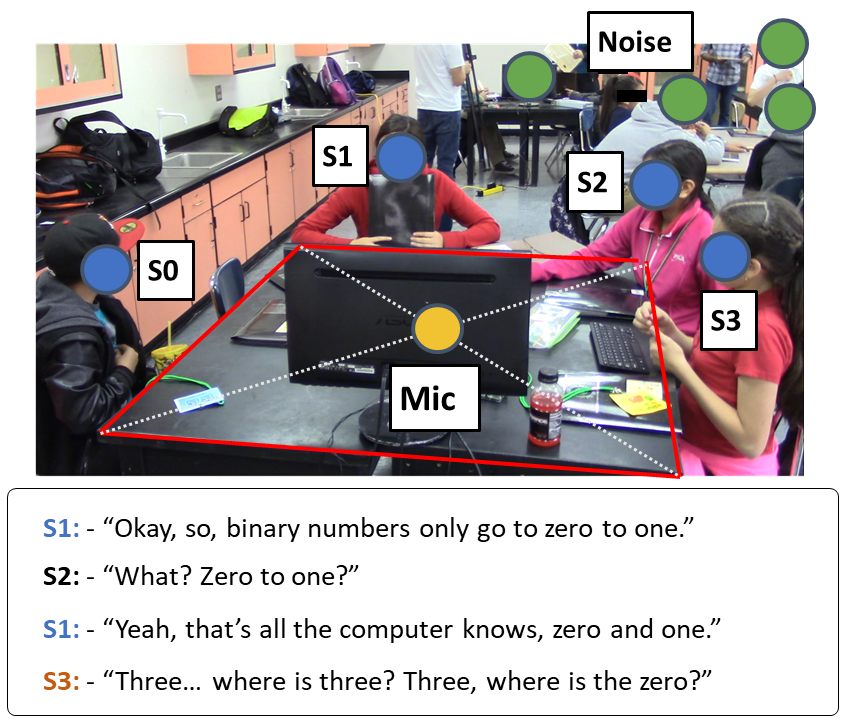}
		\caption{Example setup of a typical AOLME group interaction. Blue dots mark the speaker position and the Yellow dot is assumed to be at the center of the table (marked by red). Cross-talk is expected among speakers S0 to S3, background noise is also captured by the microphone (green dots in the back). Under the picture, we depict a sample of a transcript from the current session. Keywords can be identified like "zero", "one", "computer" and "three"	. }
		\label{fig:pyroom_intro}
	\end{figure} 
	
	We present an example in Figure \ref{fig:pyroom_intro}.
	In this example, a small group of students is sitting around 
	the table, using the keyboard to program the Raspberry Pi.
	The video has been recorded as part of 
	the Advancing Out-of-School Learning in Mathematics and Engineering (AOLME) after-school program \cite{aolme}.
	The speech recognition problem requires that we recognize what
	each of the students is saying as shown
	in the transcription of Fig. \ref{fig:pyroom_intro}.    
	More specifically, the speaker geometry requires that
	we identify the 3D locations of the speakers ($S_0, S_1, S_2, S_3$) with respect to the omnidirectional microphone placed on the center of the table.
	In Fig. \ref{fig:pyroom_intro}, we also see several other speakers talking in the background (refer to green dots).
	The students speak in both Spanish and English.
	
	Student speech recognition in this environment is very challenging
	due to cross-talk, background noise, and the use of multiple languages.
	Current deep learning systems are hence ineffective in such environments.
	To address the issue, we use the estimated 3D speaker geometry and 
	video audio transcriptions to generate a large, acoustic model based
	audio dataset that can be used to train a bilingual speech recognition
	system for this collaborative learning environment.
	As we demonstrate in this paper, although we train on synthetic
	datasets, we are still able to match and slightly exceed
	state-of-the-art systems.
	
	% prior work done in aolme.
	The current paper significantly extends our 
	previous research on analyzing such videos.
	More specifically, prior research has been focused
	on face and back of the head detection in \cite{wenjingshi}, 
	\cite{shi2}, \cite{shi3}, \cite{shiCAIP} and \cite{shiAsilomar},
		face recognition was also targeted in \cite{phuongCAIP}.
	Furthermore, authors in 
	\cite{abby1} provided an early approach to context-based activity detection using deep learning.
	The research on video activity detection was significantly extended in \cite{vjAsilomar}.
	The object detection system developed by \cite{sravani}
		will be the baseline system for estimating 3D speaker geometry from the 
		AOLME videos.	
    For completeness, we will also explain the approach in \cite{sravani} in our methodology.		
	
	The paper uses video object detection and projective geometry to locate the 3D speaker geometry from still video frames.
	The 3D speaker geometry is input to Pyroomacoustics (\cite{pyroom})
	to simulate how the speakers will be recorded by the omnidirectional microphone located on the center of the table.
	We use the audio transcriptions with the AWS text-to-speech
	system to generate the ground truth audio datasets for
	training our speech recognition system.
	The proposed approach obtained a 27.92\% recognition rate 
	on Spanish words that was slightly better than 
	Google Speech-to-text \cite{google} at 26.12\%.
	In addition, the Bilingual Keyword Classifier obtained an average of 38\% sensitivity on Spanish Keywords.
	
	% summary of the sections of the paper
	The rest of the paper is organized as follows.
	We define the 3D speaker geometry problem in section \ref{sec:cr}. 
	We then describe the underlying methods in section \ref{sec:methodology}.
	Results are given in section \ref{sec:results}.
	We then provide concluding remarks in section \ref{sec:conclusions}.
	
	\section{3D Speaker Geometry Estimation}\label{sec:cr} 
	We use projective geometry to estimate 
	3D coordinates from still image frames.
	Our basic assumption is to use cross-ratios along the projections of 3D lines to estimate 3D distances.
	We begin by assuming the basic concept and showing how to apply cross-ratios to define the problem for our videos.
	
\begin{figure}[!t]
		\centering
		\includegraphics[width=0.7\textwidth]{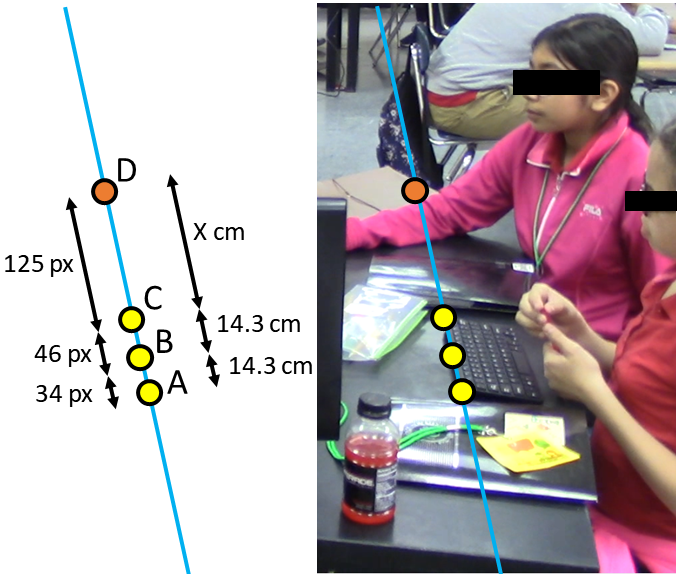}
		\caption{Physical distance estimation using cross-ratios.} 
		\label{fig:cr}
	\end{figure}

	We illustrate the concept of cross-ratios in 
	Fig. \ref{fig:cr} \cite{geometry}.         
	The basic assumption is that we know the actual physical
	distances between three consecutive, co-linear points $A, B, C$.
	In our example, let these distances be $AB$ and $BC$.
	Then, to estimate the distance to another point $D$ along
	the same line, we use the cross-ratio $R$ defined by
	\cite{geometry}:
	\begin{equation}
		R = \frac{AC}{CB}\bigg/\frac{AD}{DB} 
		= \frac{AC\cdot BD}{BC \cdot AD} 
		= \frac{(AB+BC)\cdot (BC+CD)}{BC\cdot (AB+BC+CD)} 
		\label{eq:cr}
	\end{equation}
	where $CD$ is the physical distance to be estimated.
	To estimate $CD$ from equation (\ref{eq:cr}), we first estimate the
	ratio $R$ using pixel ratios of $AD/DB$.
	Then, we substitute the value for $R$ and solve for $CD$.
	
	To estimate the 3D locations of the speakers
	using cross-ratios, we will first need to
	estimate distances along 3D planes where
	our colinear points lie.
	In our example of Fig. \ref{fig:cr},
	we assume that we know the physical dimensions
	of the keyboard (given as distance $AC$).
	Then, we estimate the midpoint $B$ of the keyboard.
	We then assume that the keyboard is parallel to the sides
	of the table (1 to 2 or 3 to 4), and 
	estimate the distance $CD$ to the edge
	of the table using cross-ratios.
	Unfortunately, we cannot use the side of the keyboard
	to estimate the width of table that is 
	depicted as a near-horizontal line in Fig. \ref{eq:cr}.
	This is because the keyboard side, compared against the table
	width is too small, and estimation can be very inaccurate.   
	
\begin{figure}[bt]
		\centering
		(a)\includegraphics[width=0.8\textwidth]{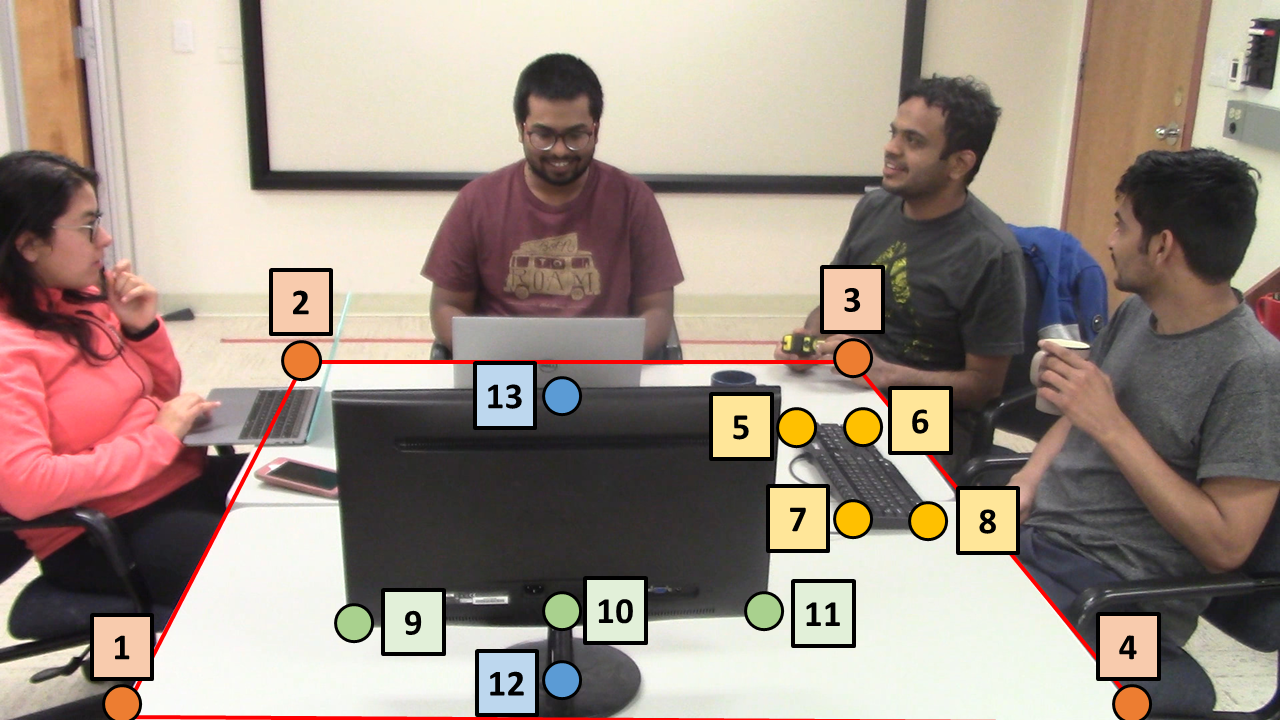}
		\\
		(b)\includegraphics[width=0.8\textwidth]{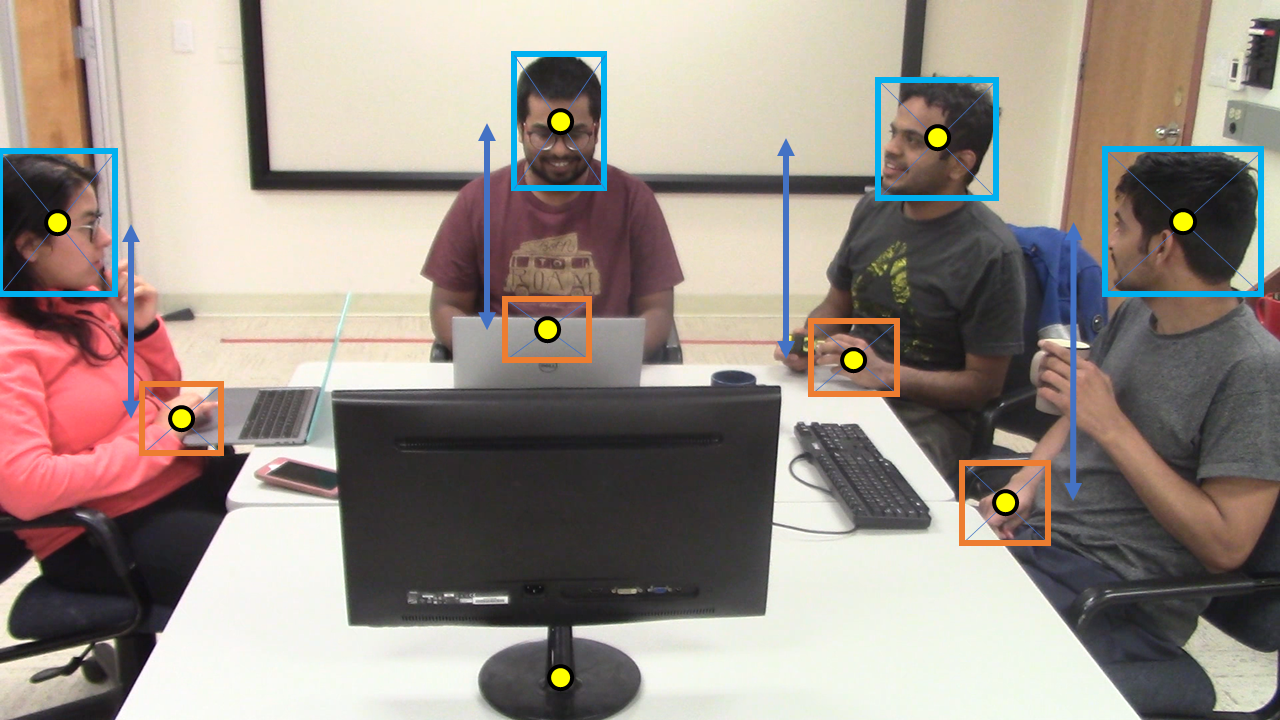}
		\caption{Speaker geometry estimation setup.} 
		\label{fig:geom}
	\end{figure}

	We define all of the points that are needed to estimate
	the 3D speaker geometry in Fig. \ref{fig:geom}.
	Here, we estimate
	all physical distances along with the table
	defined by points $1, 2, 3, 4$
	using cross-ratios.
	The basic idea is to define a 2D grid on the 
	the table that is defined through the intersection of 
	lines parallel to the keyboard (points $5, 6, 7, 8$) and the computer
	monitor ($9, 10, 11, 12, 13$).
	Here, we assume monitor points $8, 9, 10$ lie on the table to eliminate the need to map these points to the table surface.
	These lines are also assumed to be parallel to the corresponding sides of the table.
	
	Since the table is not always fully visible,
	we also extend the estimated depth
	of the table (points 1 to 2) by 5\% to account for mild occlusion.
	Here, we note that the size of the table is needed because we
	assume that the microphone is located in the center of the table.    
	
	Similarly, we define 3D planes associated with each speaker
	(assumed to be about 4 inches away from the table edge)
	and assume that the mouths and hands lie on the same
	3D plane that is orthogonal to the table.
	In terms of object recognition, we require hand detection,
	head detection as depicted in Figure \ref{fig:geom}(b).
	
	We refer to \cite{geometry2} as a base for the assumptions at 
		building the system of projections of parallel lines. 
		We plan to test at the real scenario from AOLME videos (around 1000 hours).

	\section{Methodology}\label{sec:methodology}
\begin{figure}[!t]
		\centering
		\includegraphics[width=\textwidth]{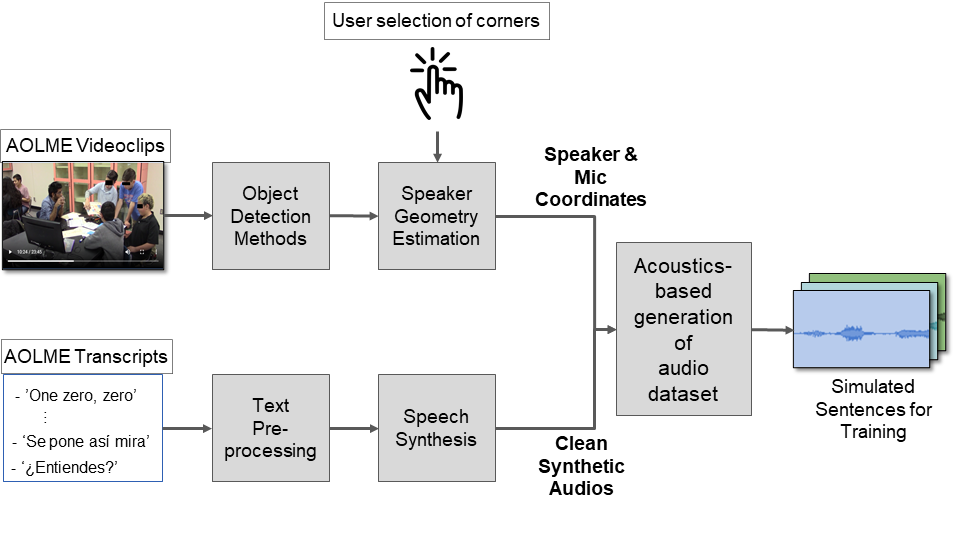}\\
		(a) Acoustics-based dataset generator based on 3D speaker geometry.\\[0.1in]
		\includegraphics[width=\textwidth]{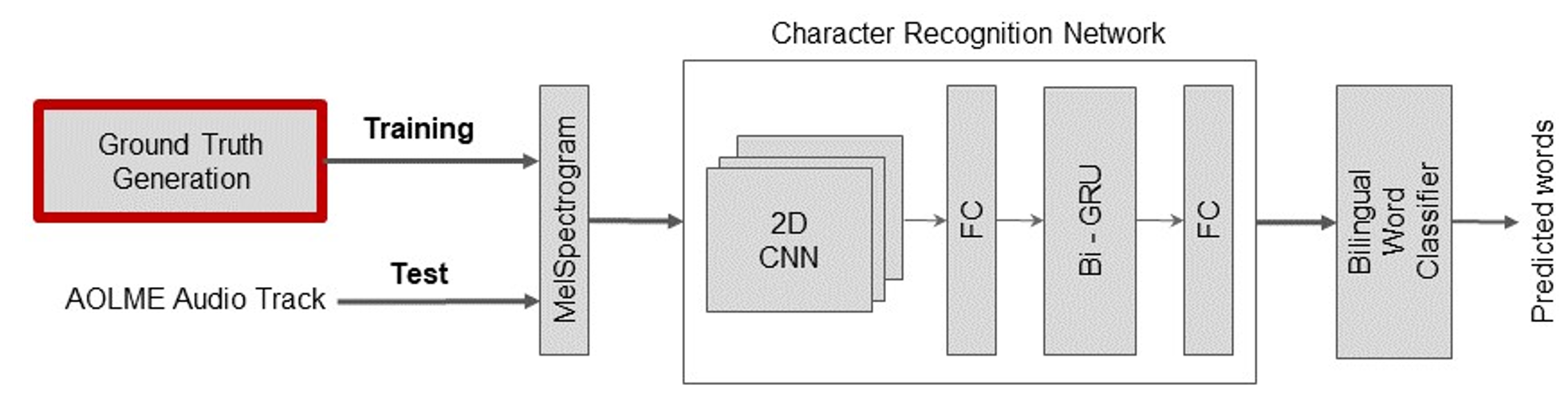}\\
		(b) Speech recognition system. \\
		\caption{Bilingual speech recognition system using 3D speaker geometry
			estimated from the video dataset.} 
		\label{fig:system}
	\end{figure}

	We summarize our methodology in Figure \ref{fig:system}.
	Our 3D speaker geometry estimation requires detection of keyboard, hands and monitor. We based the detection on \cite{sravani} with added post processing to detect necessary features to establish 3D geometry.
	We provide more details on our object detection methods in section
	\ref{sec:objrec}.
	
	Through the use of an interactive system,
	the users select specific frames, select the table corners and corners from
	the detected keyboard and monitor.
	Then, out system uses cross-ratios
	to reconstruct the 3D speaker geometry
	as summarized in section \ref{sec:cr}.
	As shown in the bottom branch of Figure \ref{fig:system}(a),
	the AOLME transcripts are pre-processed to serve as input
	to the speech synthesis module.
	We then use the reconstructed 3D geometry and the synthesized dialogues
	to provide an acoustic-based generation of the audio dataset.
	We input the 3D speaker and microphone geometry, and synthesized speech into
	our acoustic simulation framework based 
	on Pyroomacoustics \cite{pyroom}.
	The result is the acoustics-based simulated dataset
	for training our bilingual speech recognition system.
	
	The speech recognition system is shown in Figure \ref{fig:system}(b).
	The system is trained using the generated audio dataset.
	We provide more details of our speech recognition system
	in section \ref{sec:sprec}.
	
	\subsection{Object Detection}\label{sec:objrec}
	As shown in Fig. \ref{fig:geom}, we require 
	detection of the keyboard and monitor in order
	to estimate the location of the speakers with respect
	to the table.
	Furthermore, to estimate the 3D locations of the speaker's mouths,
	we also assume that their hands and mouths are on the same 3D plane
	and further require hand and head detection.   
	Here, we are only interested in hands that are located near
	the table as shown in Fig. \ref{fig:geom}.
	
	% object detection describe in a paragraph
	We next summarize the methods that we will use to detect each object.
	For head detection, we use the latest version of YOLO \cite{yolov5} 
	pre-trained on the crowd human data set for head detection \cite{shao2018crowdhuman}.
	To restrict head detection within the current student group,
	we use a minimum area threshold that successfully rejects smaller faces of people outside the group. 
	For detecting hands, monitors, and keyboards,
	  we use faster R-CNN pre-trained on the COCO dataset.
	The results of faster R-CNN 
	  are post-processed using clustering, time-projections (adding detections through time),
	  and small area removal to remove distant hands
	  (see 	\cite{sravani} for details).
	Among the hand detections, we then manually select hands that lie on the table.
    Furthermore, we manually select the edges of the Table, the monitor, and the keyboard.

	% Training data 
	We assume that we can learn the scales, number of pixels per inch for each speaker
	  using manual measurements during training.
	Later, we will look at estimating the scales for each image.
	Here, we note that our assumption is very restrictive.
	It does not account for strong scale variations when the speakers move to new positions
	   not reflected in the training set.
	
	\subsection{Speech recognition system}\label{sec:sprec} 
	We summarize the speech recognition system in Fig. \ref{fig:system}(b).
	The acoustic-based generated dataset is used to train 
	a phoneme-based recognition network composed
	of a 2D CNN (a single layer of 8 filters of size 3$\times$3 with stride=2) 
	processing Mel-spectrograms, a two-layer bi-directional GRUs with 64 units
	per layer, and a fully connected layer with an output for each phoneme.
	The system generates a sequence of phonemes characters
	that are post-processed by a bilingual word classifier
	based on minimum distance.
	
	\section{Results}\label{sec:results}
	% A summary of what kind of results we are describing
	We first summarize results from 3D speaker geometry estimation
	using a baseline approach and our proposed methods.
	We then summarize our results for 
	speech recognition system using the 3D speaker geometry.
	
	% The Baseline
	We define a baseline approach that does not require
	projective geometry or any object detection method.
	Assuming the keyboard and table corners are given,
	we assume that speakers sit around the table,
	equidistant from each other.
	
\begin{figure}[!b]
		\centering
		\includegraphics[width=0.8\textwidth]{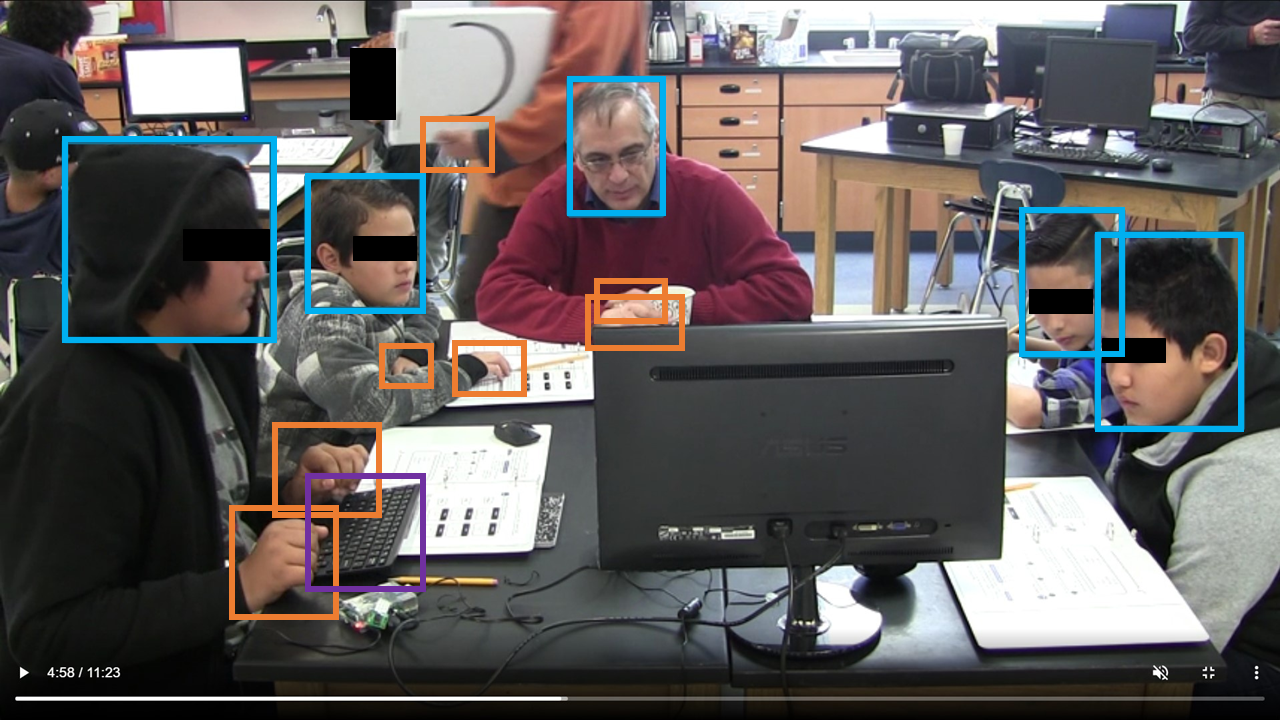}
		\caption{Object detection for 3D speaker geometry estimation.
			We use blue bounding boxes for head detection,
			orange bounding boxes for hand detection, and
			purple bounding boxes for keyboard detection.} 
		\label{fig:objdetect}
	\end{figure}

	Our proposed approach performed significantly better.
	We present a summary of our estimates for Fig. \ref{fig:geom}
	in Table \ref{tab:rad}.
	Our error ranges from 7\% to 16\%.
	The largest source of error comes from our estimation of the scale for each speaker (number of pixels per inch).
	As mentioned earlier, in future work, we will work on estimating the scale directly from each image.
	Overall, our interactive system gave
	a reduced error of 10.84\% compared to 33.12\%
	for the baseline method.
	In terms of the AOLME dataset,
	we present an example of object detection in  
	Fig. \ref{fig:objdetect}.
	Overall we note that our proposed approach required
	the combination of different object detections from different
	video frames to establish the 3D speaker geometry.

	\begin{table}[!t]
		\centering
		\caption{Results for 3D speaker geometry estimation.
			The error is given as a percentage of the distance
			to the microphone. All distances are given in inches.
		}
		\label{tab:rad}
		\setlength{\tabcolsep}{8pt}
		\begin{tabular}{llllll}
			\toprule
			\multirow{2}{*}{\textbf{Speaker}} & \multirow{2}{*}{\textbf{Ground Truth}} & \multicolumn{2}{l}{\textbf{Our Method}} & \multicolumn{2}{l}{\textbf{Baseline}} \\
			&                                        & Estimation            & Error           & Estimation           & Error          \\
			\midrule
			S0                         & 36.70                                  & 34.16                 & 6.92 \%            & 19.74                & 46.21 \%          \\
			S1                         & 35.59                                  & 41.27                & 15.96 \%           & 24.32                & 31.67 \%          \\
			S2                         & 42.12                                  & 43.88               & 4.18 \%            & 27.79                & 34.02 \%          \\
			S3                         & 34.99                                  & 29.29                & 16.29 \%          & 27.79                & 20.58 \%         \\
			\midrule
			Average 			& 37.35  & 37.15 & 10.84 \% & 24.91 & 33.12 \% \\
			\bottomrule	
		\end{tabular}
	\end{table}

	% audio results
	The output of 3D speaker geometry system is the complex simulated audio dataset, used to train the speech recognition system.
	The training dataset was generated 
	using audio transcriptions of 720 minutes
	extracted from 54 video sessions,
	and a typical AOLME 
	3D speaker geometry.
	For testing, we selected 517 sentences from
	unseen video sessions.
	We then assessed the character error rate
	for recognizing the 517 sentences.
	For this test, our proposed approach
	gave an accuracy of 
	27.92 \% compared to 26.12\% by Google speech-to-text.
	
	% Results from the Bilingual Word Classifier v2
	We also present comparative results for 
	the recognition of 9 Spanish keywords
	that were used in the number representations
	lessons.
	We summarize our results in terms of  
	sensitivity and specificity as given in Table \ref{tab:audiometrics}.
	From the results, it is clear that Google Speech-to-text
	fails to detect any instances of tres, cuatro, and cero.
	Overall, Google Speech-to-text is insensitive to the target
	keywords, it is prone to discard noisy samples as 'Others'.  
	By comparison, our proposed
	method is much better at detecting our targeted keywords
	because it will try to classify even the noisy samples.
	On average,
	sensitivity improved from 24\% to 38\% for our proposed approach.
	On the other hand, 
	specificity remained high for both methods (90\% to 92\%).       
	
	Our proposed approach produces more false positives and
	fewer false negatives than Google Speech-to-text.
	Hence, in terms of using our method, we note
	that the users would have to reject our false positive detections.
	On the other hand, Google Speech-to-text requires noise-free
	examples and fails to detect important AOLME type keywords (e.g., tres,
	cuatro, and cero).
	
	\begin{table}[bt]
		\centering
		\caption{Keyword recognition results. Here, we note that our system
			does not recognize accents.} 
		\label{tab:audiometrics}
		\setlength{\tabcolsep}{8pt}
		\begin{tabular}{lllll}
			\toprule
			& \multicolumn{2}{l}{\textbf{Our system}}     & \multicolumn{2}{l}{\textbf{Google Speech-to-Text}} \\		
			\textbf{Keywords} & \textbf{Sensitivity} & \textbf{Specificity} & \textbf{Sensitivity}     & \textbf{Specificity}    \\
			\midrule
			uno         & 0.50                 & 0.95                 & 0.13                     & 1.00                    \\
			dos         & 0.24                 & 0.91                 & 0.06                     & 1.00                    \\
			tres        & 0.63                 & 0.92                 & 0.00                     & 1.00                    \\
			cuatro      & 0.30                 & 0.99                 & 0.00                     & 1.00                    \\
			cinco       & 0.25                 & 0.99                 & 0.23                     & 1.00                    \\
			cero       & 0.36                 & 0.93                 & 0.00                     & 1.00                    \\
			computadora & 0.25                 & 0.99                 & 0.25                     & 1.00                    \\
			numero      & 0.27                 & 0.97                 & 0.45                     & 1.00                    \\
			Others      & 0.65                 & 0.67                 & 1.00                     & 0.13                    \\
			\midrule
			Average     & 0.38                 & 0.92                 & 0.24                     & 0.90        				\\
			\bottomrule           
		\end{tabular}
	\end{table}
	
	\section{Conclusions and Future Work}\label{sec:conclusions}
		
	We presented an interactive system
	for estimating 3D speaker geometries
	from a single-camera video recording.
	We then used a typical 3D speaker geometry based on AOLME videos
	to generate a complex, acoustics-based, simulated dataset
	based on 11.66 hours of audio dataset.
	Then, when tested on actual audio datasets,
	the proposed system slightly outperformed
	Google Speech-to-text. 
	% add educational researchers
	Ultimately, the detection of meaningful keywords
	can be used by educational researchers to identify
	moments of interest for further analysis.
	
	For future work, we are currently
	developing multi-objective optimization methods
	for improving our sensitivity while maintaining high specificity.

	\bibliographystyle{splncs04}
	\bibliography{cv_references}
	
\end{document}